## Highlights

- Mesoporous glasses containing cerium ions ($Ce^{3+}/Ce^{4+}$) were synthesized by EISA method.

- Beads were shaped by ionic cross-link method, using MBG-Ce powders and alginate.

- Ce ions increased catalase mimetic activity able to prevent the oxidative stress.

- The highest cell viability was attained for $CeO_2$ contents of 1.2 and 3.6 mol-%.

- Beads of glasses with 1.2 and 3.6 $CeO_2$% were able to counteract the oxidative stress.

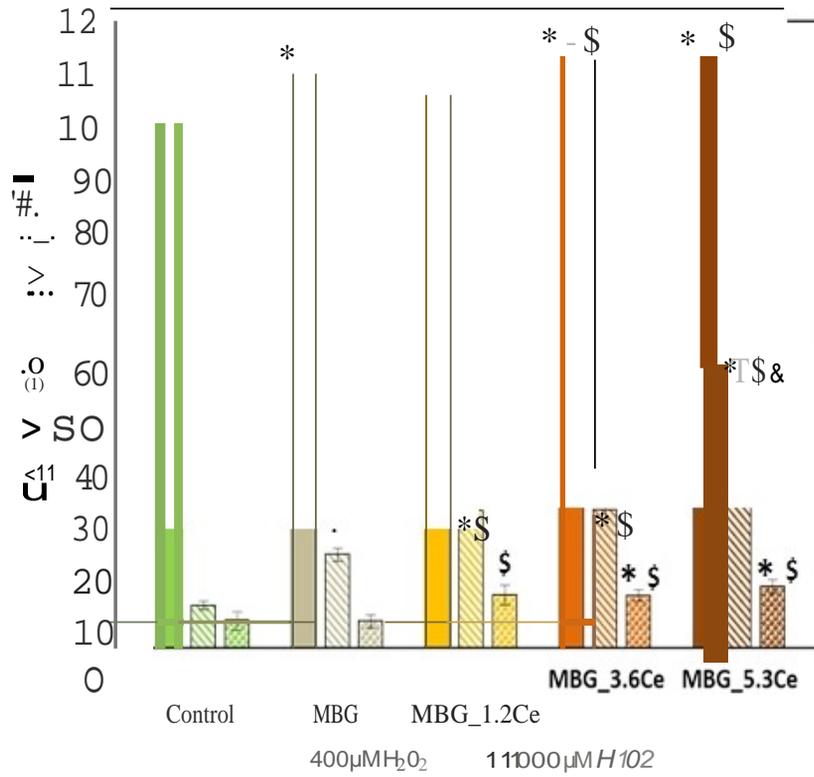
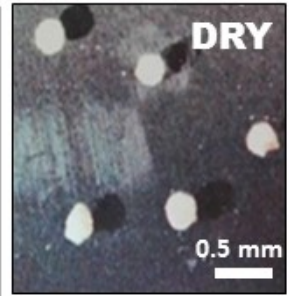
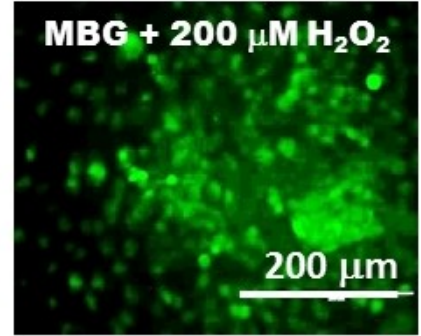
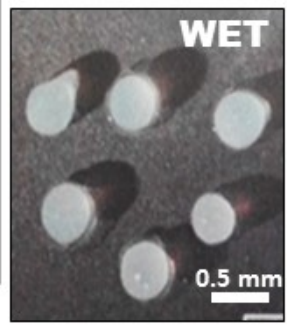
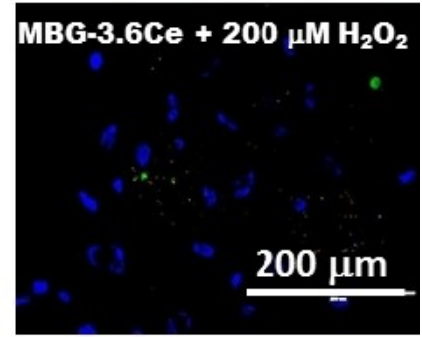

# Cerium (III) and (IV) containing mesoporous glasses/alginate beads for bone regeneration: bioactivity, biocompatibility and reactive oxygen species activity

E. Varini[a,1], S. Sánchez-Salcedo[b,c,1], G. Malavasi[a], G. Lusvardi[a], M. Vallet-Regí[b,c], A.J. Salinas[b,c*]

[a] Dpt. Chemical and Geological Sciences, University of Modena and Reggio Emilia, Via G. Campi 103, 41125 Modena, Italy.

[b] Dpt. Química en Ciencias Farmacéuticas, Facultad de Farmacia, Universidad Complutense de Madrid; Instituto de Investigación Hospital 12 de Octubre, imas12, 28040 Madrid, Spain.

[c] Networking Research Center on Bioengineering, Biomaterials and Nanomedicine (CIBER-BBN) Madrid, Spain.

* Correspondence at: A. J. Salinas (salinas@ucm.es)

§ These authors contributed equally to this work.

**A B S T R A C T**

A very small number of biomaterials investigated for bone regeneration was reported as able to prevent the oxidative stress. In this study beads based on alginate hydrogel and mesoporous glasses (MG) containing different amounts of cerium oxides ($Ce^{3+}/Ce^{4+}$) exhibiting antioxidant properties were investigated as a good approach to mimic the action of antioxidant enzymes in our organism. The effect of cerium contents on the bioactivity and biocompatibility of beads were investigated. Moreover, the potential capability of Ce-containing MG to prevent the oxidative stress caused by the activity of reactive oxygen species (ROS) was here investigated for the first time. The increment of cerium oxide from 1.2, to 3.6 and 5.3 mol-% decreases the surface area and porosity of MG and increases the catalase mimetic activity after 168 h. Swelling tests in different cell culture media (D- and α-MEM) demonstrated the rehydration capability of beads. The presence of beads with the highest Ce-contents (3.6 and 5.3 %) improved the proliferation of pre-osteoblastic cells MC3T3-C1 cells. However, the cell differentiation decreased when increased the cerium content. Lactate dehydrogenase assays showed beads are cytocompatible materials. Moreover, oxidative stress tests with $H_2O_2$ showed a better response related to cell viability and the elimination of oxidant species when increased cerium content. Beads of glasses with 1.2 and 3.6 % of $CeO_2$ are excellent candidates as bioactive scaffolds for bone regeneration capable of counteract the oxidative stress.

# 1. Introduction

Bioactive glasses (BG), discovered in 1969 by Prof. Larry Hench and clinically used since 1985, were the first synthetic materials able to bond to connective tissues and bones [1]. In contact with biological fluids, BG surfaces are coated by a bone-like nano-carbonate hydroxyapatite (n-CHA) layer that permits the interaction with the living tissues [2]. First BG were obtained by quenching of a melt, but from 1991, the interest towards the sol-gel synthesis of BG increased [3]. In 2004 the EISA method (Evaporated Induced Self-Assembly), a modification of the sol-gel method that used surfactants as structure directing agents, was proposed for the BG synthesis. That way, mesoporous bioactive glasses (MBG) exhibiting huge surface areas and pore volumes and ordered arrangements of mesopores (from 2 to 10 nm) in a very narrow pore size distribution were obtained [4]. Due to their textural properties MBG exhibited a quicker kinetics formation of n-CHA under in vitro conditions that analogous BG previously obtained [5,6]. Moreover, sol-gel and EISA methods expanded the range of glass compositions exhibiting bioactivity from a maximum $SiO_2$ content of 60 mol-% for melt glasses, to 90% for sol-gel glasses (S-GG) and MBG [7,8]. Porosity of S-GG and MBG allow the formation of a hydrated layer inside the material, where biological moieties can penetrate maintaining their structural configuration and biological activity and becoming an indistinguishable part of the host tissue. Thus, when trabecular rabbit bone was proliferated on a melt BG, a structure similar to bone was obtained, but some BG particles were still present. Instead, when S-GG were used, no residual particles were observed [9].

On the other hand, the implantation of biomaterials could have as consequence the so-called surgical stress response: a physiological mechanism that involves the activation of inflammatory, endocrine, metabolic and immunologic mediators [10,11]. Among the effect of this mechanism, there is the occurrence of oxidative stress, with production of reactive oxygen (ROS) or nitrogen (RNS) species that stoke the inflammation. It has been demonstrated that the administration of antioxidants results in a faster recovery after the operation. Thus, the design of biomaterials with antioxidant properties would shorten the convalescence and reduce the amount of anti-inflammatory drugs administered to patients [12]. Cerium oxide nanoparticles (CeONPs or nanoceria) have received much attention because of their excellent catalytic activities, which are derived from quick interconversion between $Ce^{4+}$ and $Ce^{3+}$ oxidation states. The cerium atom has the ability to easily and drastically adjust its electronic configuration to best fit its immediate environment. The ability of CeONPs to present the redox $Ce^{4+}/Ce^{3+}$ interconversion is due to a specific characteristic of the surface of nanoparticles: they exhibit oxygen vacancies in the lattice structure; these arise through loss of oxygen atoms, alternating between $CeO_2$ and $CeO_{2-x}$ during redox reactions. Nanoceria was recently found to have multi-enzyme mimetic properties, including superoxide dismutase (SOD), catalase and oxidase, which produces various biological effects, such as being potentially antioxidant towards almost all noxious intracellular reactive oxygen species [13,14]. Nanoceria has emerged as a fascinating and lucrative material in biological fields such as bioanalysis, biomedicine, drug delivery and bio-scaffolding [15].

Recently, we developed promising BG based on Hench´s 45S5 Bioglass® modified by $Ce^{4+}/Ce^{3+}$ ions (additions of $CeO_2$ during the synthesis of glasses) to join the ability of material to promote the bond with hard tissues (bioactivity, i.e. n-CHA formation) and simultaneously to show enzyme-like activities (Catalase and SOD) [16–18]. BG play a critical role in the success of hard tissue engineering, and the selection of the appropriate material can have a profound impact on the quality of newly formed bone tissue. A major challenge facing the field of tissue engineering is the development of materials capable of promoting the desired cellular and tissue behavior. Given that few biomaterials possess all the necessary characteristics to perform ideally, engineers and clinicians alike have pursued the development of hybrid or composite biomaterials to synergize the beneficial properties of multiple materials into a superior matrix. The combination of natural and synthetic polymers with various other materials has demonstrated the ability to enhance cellular interaction, encourage integration into host tissue, and provide tunable materials that can promote the formation of bone, vascular, and neural tissues. The ideal tissue engineered construct is a porous interconnected structure that allows cells to migrate and function within its confines (osteoconductive), provides factors that stimulate the proliferation and differentiation of progenitor or osteogenic cells (osteoinductive), and is capable of assimilating into the surrounding tissue (osteointegrative), eliminating a potential infection [19–21].

Alginate is a naturally occurring polymer (polysaccharide) typically obtained from brown seaweed and has been extensively investigated and used for many biomedical applications, due to its biocompatibility, low toxicity, relatively low cost, and mild gelation by addition of divalent cations such as $Ca^{2+}$. Alginate is of particular interest for a wide range of applications as a biomaterial and especially as a supporting matrix or delivery system for tissue repair and regeneration. Due to its outstanding properties, alginate was widely used in a variety of biomedical applications including tissue engineering, drug delivery and formulations preventing gastric reflux [22]. Alginate is readily processable for applicable three-dimensional scaffolding materials such as hydrogels, microspheres, microcapsules, sponges, foams and fibers. Alginate hydrogels can be prepared by various cross-linking methods [23,24], and their structural similarity to extracellular matrices of living tissues allows wide applications in wound healing, bone regeneration and cartilage repair [25].

The creation of composites is a biomimetic approach, as bone can be viewed as a composite of collagen, n-CHA and in minor quantity water and small amounts of other organics. Not surprisingly, improvement in regeneration has been observed in composite constructs mimicking the composition and structure of bone phases [26]. The material degradation will make room for the growth of new tissue, allow for the integration of any delivered cells into the surrounding tissues, and release incorporated bioactive species [27]. The introduction of BG powders into the cross-linked alginate hydrogel is intended to retard the glass degradation into the body and to avoid the direct contact with cells, because these glasses release their alkaline earth cations inducing a pH augment that brings to cells' death [28,29]. In the literature there are a very small number of scaffolds designed for bone tissue engineering that are able to prevent the oxidative stress. Therefore, introducing glasses with different amounts of cerium into the alginate hydrogel, to check if the resultant composite exhibits antioxidant properties mimicking the action of the enzymes in our organism, is a subject of interest. Therefore, the aim of this work is to develop biocompatible Ce-MBG/alginate beads able to promote a fast degradation of $H_2O_2$ to reduce the oxidative stress.

Studies previously reported in literature are devoted to developing biomaterials (ceramics or polymers) capable of releasing molecules, ions or nanoparticles that act as antioxidants [30–32]. In our case, we intend to develop materials capable of carrying out their action by promoting the antioxidant activity at the interface between the biomaterial itself and the physiological environment where they were implanted.

In this study, four mesoporous bioactive glasses first Ce-free (MBG, $80SiO_2–15CaO–5P_2O_5$ mol-%), the other three modified with 1.2, 3.6 and 5.3 mol-% of $CeO_2$ (Ce-MBG), were synthetized. Moreover, beads were processed by ionic cross-linking of the BG powders and alginate. The experimental characterization of samples will allow to determine the effect of alginate and cerium ions on the morphological and textural properties, bioactivity and catalase-like activity of samples. The effect of cerium on the materials biocompatibility was also investigated and, the ability of Ce-MBG to prevent the oxidative stress caused by ROS was here investigated for the first time.

## 2. Materials and Methods

### 2.1 Synthesis of mesoporous bioactive glasses (MBG)

The MBG synthesis was performed via sol gel method by using the evaporation induced self-assembly process (EISA) [4,8] with the non-ionic surfactant Pluronic® P123 as a structure-directing agent (SDA). P123 is an amphiphilic triblock copolymer. In this work, a series of MBG with a starting composition of $80\%SiO_2–15\%CaO–5\%P_2O_5$ were synthesized. The glasses have been added with increasing amounts of $CeO_2$. The theoretical molar compositions of the samples are reported in Table 1 and named hereafter MBG with the relative mol-% of cerium oxide. Tetraethyl orthosilicate (TEOS) (98%, Sigma Aldrich), triethyl phosphate (TEP) (99%, Sigma Aldrich), calcium nitrate tetrahydrate $Ca(NO_3)_2 \cdot 4H_2O$ (99%, Sigma Aldrich) and cerium nitrate hexahydrate $Ce(NO_3)_3 \cdot 6H_2O$ were used as $SiO_2$, $P_2O_5$, CaO and cerium oxides sources, respectively.

Pluronic® P123 (4.5 g) was dissolved by magnetic stirring (~ 1 h) in 85 mL of ethanol containing 1.2 mL of 10% HCl solution; afterward the appropriate amounts of TEOS, TEP, $Ca(NO_3)_2 \cdot 4H_2O$ and $Ce(NO_3)_3 \cdot 6H_2O$ were added, under continuous stirring in 3 h intervals at room temperature. The resulting sols were kept under stirring overnight and then cast in Petri dishes (9 cm in diameter) to undergo the EISA process. Colorless solutions were allowed to evaporate at 25°C for several days. Successively, the dried gels were removed as homogeneous and transparent membranes and heated at 700 °C for 3 h under an air atmosphere to remove the surfactant and nitrate groups, as well as to stabilize the resultant mesoporous glasses. Finally, MBG were gently milled and sieved at a mean dimension lower than 250 μm, giving rise to the P-MBG, P-MBG_1.2Ce, P-MBG_3.6Ce and P-MBG_5.3Ce powders.

## 2.2 Preparation of alginate/glass beads

The alginate beads were synthesized by ionic cross-linking method (see **Scheme 1**) using a 1% w/w a solution of sodium alginate (dissolution of 0.20 g of sodium alginate in 20 mL of Milli-Q water). After the complete dissolution and continuous stirring, 0.50 g of glass powder were added (< 250 μm) to obtain a turbid solution. Transfer the mix alginate/glass in a plastic syringe without needle (30 mL) and add it drop by drop to the 0.1 M $Ca^{2+}$ solution prepared dissolving $CaCl_2$ in Milli-Q water. This procedure enables to obtain sufficiently homogeneous spherical beads. Let the obtained solution under stirring for 2 h. Beads were filtered under vacuum, washed with Milli-Q water and ethanol to facilitate their separation. Then, beads were plated in Petri dishes and let them to dry for 2 h at 60 °C. Wet and dry beads showed average dimensions of 3.9 and 2.1 mm respectively. The shape of the beads can be considered approximately spherical, giving rise to the B-MBG, B-MBG_1.2Ce, B-MBG_3.6Ce and B-MBG_5.3Ce beads.

## 2.3 Physical-Chemical Characterization of powders and beads

MBG powders (P-MBG) were characterized to investigate their chemical composition, X-ray fluorescence spectroscopy (XRF), textural properties ($N_2$ adsorption), mesoporous structure, X-ray diffraction and transmission electron microscopy (XRD and TEM) and the glass network connectivity solid state Nuclear Magnetic resonance (NMR).

*2.3.1 X-ray fluorescence spectroscopy (XRF)*. The chemical composition of the mesoporous bioactive glasses was determined by XRF measurements. The samples are prepared as a flat disk, of diameter 13 mm, of boric acid (0.20 g) with the MBGs powder (0.02 g) distributed to the entire surface of the top face of the disk. Measurements were performed in triplicate and average values with ±std. dev. are reported in the **Table 1**.

*2.3.2 Textural properties*. Surface area, pore volume and mean pore diameter of glass powders were determined by $N_2$ adsorption/desorption isotherms carried out at T ~ 77 K using Micromeritics ASAP 2020 and ASAP 2010 Surface Area and Porosity Analyzer. Adsorption data were processed by: (i) the standard Brunauer, Emmet and Teller (BET) method [33] to determine specific surface area ($S_{BET}$); (ii) the Barrett, Joyner and Halenda (BJH adsorption) method [34] to determine mesopores size distributions and (iii) single point adsorption to determine pore volume ($V_P$).

*2.3.3 Small-angle X-ray diffraction (SA-XRD)* SA-XRD was done to explore the possible mesopores order in the MBGs powder samples. SA-XRD experiments were performed in an X-ray diffractometer Panalytical X'Pert MPD with parallel optics (diffracted beam collimator) equipped with Cu Kα (wavelength 1.5418 A), 45 kV, 40 mA, Bragg-Brentano conditions. XRD patterns were collected in the °2θ range of 0.6°-6.5° with a step size of 0.02° and a counting time of 5 s per step.

*2.3.4 Transmission Electron Microscopy (TEM)*. TEM was used to visualize the mesopores structure, in particular the pores structural arrangements. To perform the imaging, the MBGs powders were dispersed in acetone in an ultrasonic bath for 2 min and then the suspension was deposited on a copper grid (φ = 2 mm). Measurements were performed with a JEM 2100HT (JEOL multipurpose analytical high-resolution transmission electron microscope). The accelerating voltage was 200 kV and the point resolution of 0.25 nm. The energy dispersive X-ray microanalysis was done using the system EDS – Oxford INCA.

*2.3.5 Solid-state nuclear magnetic resonance (NMR) spectroscopy*. NMR spectroscopy was used for exploring structural order of amorphous materials at short (≲0.3 nm) and medium (≲1 nm) ranges. The technique reveals intermediate-range structural order, such as information about connectivity among the basic building blocks in the glass networks and how more extended structural motifs may be identified [35]. Single-Pulse/Magic angle-spinning (SP/MAS) solid-state nuclear magnetic resonance measurements were performed to evaluate the local environment of Si atoms, thus elucidating the silicon species that are contained in the synthesized samples. The NMR spectra were recorded on ad Advance 400 Wide Bore spectrometer (Bruker) equipped with a solid-state probe using a 4 mm zirconia rotor and spun at 10 kHz for $^{29}$Si. Samples were spun at 10 kHz for $^{29}$Si. Spectrometer frequencies were set to 79.49 and 161.97 MHz for $^{29}$Si. Chemical shift values were referenced to tetramethylsilane for $^{29}$Si. All spectra were obtained using a proton enhanced cross polarization method, using a contact time of 1 ms. The time periods between successive accumulations were 5 s for $^{29}$Si, and the number of scans was 10 000 for all the spectra.

Alginate/glass beads (B-MBG) were characterized to compare the textural properties with those of the P-MBG powders (N$_2$ adsorption), determine the amount of glass they contained into by thermal decomposition (TGA), the $Ce^{3+}/Ce^{4+}$ ratio on the glass surface before and after the beads formation (XPS) and study the morphology structure (SEM) and their rehydration capability (swelling). Textural properties where determined as previously reported for MBG powders.

*2.3.6 Thermogravimetric analysis (TGA).* TGA was performed on the beads (alginate/glass samples B-MBG) to determine the average ratio between alginate and glass. The measurements were conducted, in triplicate, on all the MBG/alginate beads and on glass-free alginate beads as a reference, and average values are reported in the data. TGA analysis were performed up to 1000 °C using 50 mg of the powder samples with a 10 °C min$^{-1}$ heating rate. The atmosphere was air. The accuracy in the determination of mass loss was due primarily to the accuracy of the TGA measurements (0.2 mg). In fact, the difference between three replicates was less than 0.2 mg. TGA equipment was calibrated periodically using CuSO$_4$·5H$_2$O. The instrument was a LABSYS - Setaram, and the data were recorded and analyzed with a computer interfaced with and TGA equipment.

*2.3.7 X-ray Photoemission Spectroscopy (XPS).* XPS spectra were measured in an ultra-high vacuum apparatus using Al Kα photons as the exciting probe and a 125 mm mean radius hemispherical electron analyzer. We carried out an analysis of Ce 3d XPS lines of all AQ samples to evaluate the $Ce^{3+}/Ce^{4+}$ concentration ratio on their surface. The analysis of Ce 3d photoemission is based on the fitting of the spectra using five doublets, two related to $Ce^{3+}$ ions and three to $Ce^{4+}$ ions in different final states [36]. A Shirley-type background and five Voigt-shaped doublets are used to fit the spectra, using fixed energy positions and widths, chosen as close as possible to values determined by previous works [37], and with only the peak intensities as a free fitting parameter. The $Ce^{3+}/Ce^{4+}$ ratio is evaluated by calculating the ratio of the areas of the $Ce^{3+}$- and $Ce^{4+}$- related peaks in the best fit of each spectrum. The same procedure was used in several of our previous works. [16,17]

*2.3.8 Scanning Electron Microscopy (SEM)* The morphology of the beads before and after Simulated Body Fluid (SBF) soaking was examined by SEM using a JSM-6335F (JEOL) microscope operating at 20 kV.

*2.3.9 Swelling Test*s. Alginate/glass bead rehydration behavior is a fundamental property in view of potential application as scaffolds [38]. Rehydration assays were performed in Milli-Q water, SBF, Dulbecco's Modification of Eagle's Medium (D-MEM) and Minimum Essential Medium (α-MEM) to investigate and compare the swelling/degradation behavior of MBGs/alginate beads in different media. A bead of each type of bioactive glass (3 replicates) was soaked in 2 mL of each fluid in a 24 well-plate at 37°C to mimic the conditions of the cytocompatibility assays at 2, 4, 6 and 24 h. The mass measurement process continued until the swollen beads could be weighed properly. The dynamic weight change (%$W_{REHYDRARTATION}$) of the beads with respect to time, defined also as swelling ratio ($S_w$), was calculated according to the following equation:

where $W_t$ is the weight of rehydrated beads at time t and $W_d$ is the weight of initial dried beads [39]. The tests were carried out in triplicate.

$$\%W_{REHYDRATATION} = 100 \times \frac{W_t - W_d}{W_d}$$

*2.3.10 Catalase mimetic activity tests.* Catalase mimetic activity tests of powders and beads were performed to verify their ability to mimic the catalase antioxidant activity. The catalase mimetic activity tests have been performed as reported in our previous paper [16]. We have soaked the sample in a solution of hydrogen peroxide 1 M, into a polyethylene bottle, with a glass/liquid ratio of 5 mg/mL. The soakings have been carried out for 1, 2, 4, 8, 24, 48, 96 and 168 h, under continuous stirring at 37°C. After the soaking, we have filtered the solutions and we have titrated them to quantify the residual hydrogen peroxide. Three independent samples were tested and the results were expressed as mean value associate to the standard deviation (±std. dev.).

## *2.4 In vitro bioactivity assays*

These tests were performed in simulated body fluid (SBF) following the static method, in which the solution used for soaking biomaterial is never renewed throughout the experiment. SBF was prepared according to Kokubo's protocol [40]. Each powder sample (65 mg) was soaked for 5 different times (8 h, 1 d, 4 d, 7 d and 14 d) and the beads samples (99 mg) for 6 different time intervals (8 h, 1 d, 4 d, 7 d, 14 d and 28 d) in an airtight polyethylene container with 12 mL SBF at 37°C. For the beads, the decision to perform the 28 d time interval was made considering that the bioactivity could be slowed by the presence of the alginate. In this work, it has been used the same glass/solution ratio used in a previous work [41]. When the beads were soaked, the amount of beads to be tested was determined taking into account that the average content of glass in each bead was around 65%.

The polyethylene bottles containing the samples were placed in an incubator set at a temperature of

37°C. At the end of each time period, the sample was removed from the incubator and the solids were collected by filtration. The solid was immediately washed with DI water and subsequently with acetone to terminate any reaction and dry the sample quickly. Each sample was run in duplicate. The same protocol was applied to the "blank" SBF solution as reference. The experiments were carried out in triplicate. The evolution of surface layer of the samples soaked in SBF was studied by Fourier infrared spectroscopy using attenuated total reflection (FTIR-ATR) stage, Wide Angle X-ray diffraction (WA-XRD), scanning electron microscopy and energy dispersive spectroscopy (SEM-EDS) using the procedure describe above.

*2.5 In vitro biocompatibility assays*

Prior to all in vitro biocompatibility assays the beads were washed with distilled water and ethanol to remove the presence of any chlorides (synthesis residual) and sterilized under ultraviolet light for 1 hour to avoid cells contamination. Controls tissue cells in absence of samples were always carried out. All the experiments were performed in triplicate and average values are reported in the data.

In vitro biocompatibility tests were performed using mouse calvaria preosteoblastic cell line MC3T3-E1 (sub clone 4, CRL-2593; ATCC, Mannassas, VA) [42].

Cells were sub-cultured in a P75 flask (cell growth surface area 75 cm$^2$) with complete medium (10 mL) that consists on Minimum Essential Medium (MEM Alpha Medium 1X) containing penicillin/streptomycin (Antibiotic-Antimycotic (100X)), L-glutamine 200mM (100X) and 10 % fetal bovine serum (FBS) working in a humidified atmosphere of 95% air and 5% $CO_2$ at 37°C. All the reagents have been purchased by Gibco® by life technologies™.

*2.5.1 Cell proliferation: mitochondrial activity.* Cell proliferation was measured in terms of metabolic activity by means of the Alamarblue® method at 1 and 4 d of culture. The Alamarblue® method is based on the reduction of blue fluorogen (resazurin) to a red fluorescent compound (resofurin) by intracellular redox enzymes. Two different types of assays were performed: (1) direct assay, the cells were seeded directly on the material and, (2) indirect assay the materials were introduced in Transwells® (Corning Incorporated, Costar®, USA) and soaked in the complete medium the day after the cells were planted in the wells. MC3T3-E1 cells were plated in 24-well tissue culture plates (Sarstedt, Germany) and made in contact with the samples (as described above) at a seeding density of 5 x 10$^5$ cells/mL in complete medium (2 mL/well) for 1 and 4 d at 37°C in a humified atmosphere of 5% $CO_2$. The morphology of the cells was visualized by using EVOS Digital Inverted Microscope (AMG, Advanced Microscope Group). The AlamarBlue® reduction test was performed using a commercial assay and following the manufacturer's protocol (AlamarBlue® Cell Viability Reagent, Invitrogen). The reagent was mixed with complete medium without Red Phenol (ratio 1:10) and the final solution was added in each well (0.5 ml). The scaffolds were moved to new wells and the cells were exposed to resazurin solution for 4 h at 37 °C under $CO_2$ (5%) atmosphere. Then fluorescence signal was read at λem = 590 nm using a λexc = 560 nm with a fluorescence spectrometer Biotek Wave Power XS (program Gen5, version 1.00.14). The medium was renewed to continue the cell proliferation measurements at different times.

*2.5.2 Cytotoxicity assay: lactate dehydrogenase (LDH) activity.* Extracellular lactate dehydrogenase (LDH) activity was measured at 1 and 4 d of seeding in the culture medium, using 117 μl of the culture medium and 0.5 mL of the commercially available kit for quantitative determination of LDH (SPINREACT, Spain). The LDH activity was measured using a Unicam UV 500 UV–Visible spectrophotometer (Thermo Spectronic, Cambridge, UK) at 340 nm in the culture medium following the manufacturer protocol.

*2.5.3 Cell differentiation: ALP activity.* Alkaline phosphatase (ALP) activity of cells growing onto the scaffolds after 7 d of incubation was used as the key differentiation marker in assessing the expression of the osteoblast phenotype. For this purpose, MC3T3-E1 preosteoblasts (0.5 x $10^4$ cells/ml) were seeded using supplemented medium with β-glycerolphosphate (50 mg $mL^{-1}$, Sigma Chemical Company, St. Louis, MO, USA) and L-ascorbic acid (10 mM, Sigma Chemical Company, St. Louis, MO, USA). Every 3 d the medium was changed to ensure the cells the necessary sustenance. After 7 d, the medium was removed and the plate was involved in 3 cycles of freezing at -80°C for 30 min and defreezing at room temperature for the same time, to induce the lysis of the cellular membrane and the liberation of the ALP. ALP activity was measured on the basis of the hydrolysis of p-nitrophenylphosphate to p-nitrophenol. After 30 min incubation at 37 °C the reaction was stopped by the addition of 200 μl of 2N NaOH. The absorbance of the final solution was measured spectrophotometrically at 405 nm with a Unicam UV-500 UV– visible spectrophotometer (Thermo Spectronic, Cambridge, UK) and is proportional to the alkaline phosphatase activity. A blank sample, with the working reagent, was utilized as a reference.

*2.5.4 Evaluation of the biocatalytic activity in presence of hydrogen peroxide.*

*(A) Cell viability*. The assays were conducted using a colorimetric method that allow the determination of the number of viable cells in proliferation or cytotoxic assays. The measurements were made using a commercial kit, CellTiter 96® AQueous Non-Radioactive Cell Proliferation Assay (MTS) (Promega Corporation, USA). The conversion of MTS into the aqueous soluble formazan product is accomplished by dehydrogenase enzymes found in metabolically active cells. The quantity of formazan product as measured by the amount of 490 nm absorbance is directly proportional to the number of living cells in culture. The measurements were performed in a UV-visible spectrophotometer Biotek Wave Power XS (program Gen5, version 1.00.14).

*(B) ROS test.* The reagent utilized was the 2′,7′-dichlorofluorescin diacetate (DCFDA, ≥97% D6883 Sigma Aldrich); it is a cell-permeable non-fluorescent probe. 2′,7′-dichlorofluorescin diacetate is de-esterified intracellularly and turns to highly fluorescent 2′,7′-dichlorofluorescein upon oxidation (DCF). In this colorimetric assay to evaluate the production of oxidative species afterwards the introduction of the hydrogen peroxide, the measurements were made using an EVOS Digital Inverted Microscope and filter GFP (AMG, America Microscope Group).

## 3. Results and Discussion

### *3.1* **MBG materials and beads characterization**

**Table 1** includes the experimental glass powder composition. The molar percentages of $SiO_2$, $CaO$, $P_2O_5$ and $CeO_2$ were very similar to the nominal composition. The measurements were conducted considering three different points of each samples and the data are closer to the theoretical one than XRF data. **Figure S1** (Supplementary material) shows SA-XRD patterns of all the mesoporous glasses synthesized. The absence of intense and sharp maxima in the patterns of the P-MBG_3.6Ce and P-MBG_5.3Ce point out the wormlike structure of these materials as showed in TEM images reported in **Figure S2** (Supplementary material). In order to control the surface modification induced by the composite (glass/alginate) formation an XPS analysis was performed on the MBG_5.3Ce system before (P-MBG_5.3Ce) and after (B- MBG_5.3Ce) bead formation. The result was reported in **Figure S3 (a)** (see Supplementary material): the spectra showed that the $Ce^{3+}/Ce^{4+}$ ratio (R) still remain the same before and after the composite glass/alginate formation confirming that the interaction between the glass powder and the polymer do not affected the chemical composition of glass surface. Moreover, in **Figure S3 (b)** the FTIR spectra of P-MBG_5.3Ce, B-MBG_5.3Ce and B-Alg samples are reported. The differences between the sample before (P-MBG_5.3Ce) and after (B-MBG_5.3Ce) the interaction of glass with the alginate matrix are due to the presence of alginate (especially band at around 1630 and 1420 cm$^{-1}$); the vibrations due to the silicate network do not change significantly before and after the formation of beads confirming that the interaction of alginate matrix with the glass sample does not modify the glass structure.

SP/MAS NMR $^{29}$Si results were utilized to evaluate the environment of Si atoms, also considering the effect of the cerium contained in the glasses on the chemical shift and area of the peaks. **Figure 1** included the $^{29}$Si NMR spectra for all the BG synthesized. All the spectra displayed resonances at around -90, -104 and -111 ppm that represented $Q^2$ [(NBO)$_2$-Si-(OSi)$_2$], $Q^3$ [(NBO)-Si-(OSi)$_3$] and $Q^4$ [Si-(OSi)$_4$] silicon sites, respectively (NBO = non-bonding oxygen) [42]. $Q^n$ speciations of P-MBG were in agreement with data obtained in previous work [42]. The value of $Q^4$ peak area was the highest compared to the Ce-containing samples: there is a predominance of [Si-(OSi)$_4$] unit, so this glass shows a high degree of polymerization and connectivity. The diminution of the network connectivity (decrease of $Q^4$ and increase of $Q^2$ and $Q^3$ areas), in P-MBG_1.2Ce, P-MBG_3.6Ce and P-MBG_5.3Ce, respect to P-MBG was evident. This trend is in agreement with the role of cerium in the glass network; the field strength of the Ce$^{3+}$ is 0.48 (calculated based on Ce-O = 2.49 A with Ce in dodecahedral coordination, CN = 8) and it is usually considered as an intermediate ion with a major tendency to behave like a glass modifier [43].

In P-MBG_1.2Ce and P-MBG_3.6Ce, the cerium ions are network modifiers producing a depolymerization of the glass structure. In P-MBG_5.3Ce, there was a small increase of $Q^3$ and a decrease of $Q^2$; while, the value $Q^4$ is the almost same of the other Ce-containing samples (see **Figure 1**). This last result can be explain based on structural behavior of cerium ions. In fact, in previous works performed of melt-quenched potential bioactive Ce-containing glasses [16,17] cerium ions, especially at high concentration, can segregate and form clusters near the phosphate units. In this way, the silicate matrix becomes depleted of cerium ions (decrease the amount of network modifiers), therefore it increases the degree of polymerization.

**Figure 2** shows SEM images of a beads containing MBGs powders. From the micrographs it can be noticed the pieces of the glass powders that are incorporated and held together in a polymeric matrix of alginate. **Figure 2** left shows the morphology of B-MBG and B-MBG_3.6Ce beads; the surface is characterized by a partially homogeneous distribution of the glass powders and by the presence of volume defects like fractures, probably formed during the drying process at 60°C. The micrographs showed at the right of **Figure 2** illustrate the cross-section of B-MBG_1.2Ce and B-MBG_5.3Ce. In the fractured beads, it can be observed the granules of bioactive glasses, randomly distributed and supported in the polymeric matrix, creating a porous structure.

N$_2$ adsorption/desorption measurements were performed to evaluate and compare textural properties (surface area, pore diameter and pore volume) of the potential bioactive glass powders and beads. In **Table 2** are represented the textural parameters of the samples. Specific surface area ($S_{BET}$), pore diameter ($D_P$) and pore volume ($V_P$) decrease from powders to beads and with the increase of the cerium content. The decrease of textural properties of the beads agrees with the presence of the polymer that covers the glass surface, resulting in lower surface areas and pore volumes, maintaining the mesopores structure. However, the beads maintain a significant surface area and mean pore dimensions useful to improve the in vitro formation rate of n-CHA [44]. The effect of alginate on the textural properties is very similar to that of the cerium. In fact, increasing the CeO$_2$ into the glass the $S_{BET}$, mean pore diameter and $V_P$ decrease. The thermal degradation curve of beads and the swelling behavior determined by gravimetric analysis were included in Supplementary material in **Figures S4** and **S5**, respectively.

## *3.2 Catalase mimetic activity assay*

**Table 3** reports the data collected in the catalase mimetic activity tests. The degradation of $H_2O_2$ after soaking the samples for different times (1, 2, 4, 8, 24, 48, 96 and 168 h) was checked. The stability of 1 M $H_2O_2$ solution was also tested: in the time range of catalase mimetic activity tests, the solution is very stable because the $H_2O_2$ concentration remains substantially the same over the whole period (0.98 M after 168 h of incubation). The $H_2O_2$ molar concentration decreases as a function of time for all the samples (powders and beads) and the degradation of $H_2O_2$ is faster as a function of cerium content in the glass composition as previously detected [16–18]. In particular for P-MBG_5.3, B-MBG_5.3 and B-MBG_3.6 samples the $H_2O_2$ decomposition was complete after 168 h.

The alginate matrix seems that did not substantially modify the catalase mimetic activity, at most the data suggested a slightly increment of catalase mimetic activity of the beads (B-MBGs) with respect to the correspondent glasses (P-MBGs). This could be explained by a better dispersion of the glass powders in the alginate matrix when the sample were soaked in $H_2O_2$ solution. In fact, during the tests the glass powders tend to settle and compact on the bottom of the polyethylene container; while the beads after their deposition on the bottom of the bottle formed a network richer in cavities accessible from the $H_2O_2$ solution.

To evaluate the antioxidant activity regarding ROS, we tested the superoxide dismutase (SOD) mimetic activity for the systems studied. For reasons of brevity, the results are not reported here, however we can say that all glass/alginate systems containing cerium ions show SOD mimetic activity. A detailed study about this enzymatic-like activity is in progress.

## *3.3 Bioactivity assays*

The FTIR spectra after the bioactivity assays are reported in **Figure S6** in Supplementary material. The bands at 564 and 605 cm$^{-1}$, usually assigned to crystalline calcium phosphate hydroxyapatite, was clearly evident after 1 d for P-MBG glass, while the doublet was also present in Ce-containing samples, but less evident. This could be due to the fact that the cerium usually formed insoluble cluster with $PO_4^{3-}$, so the phosphate ions could not participate to the formation of the crystalline layer, causing a decrease the bioactive property of the material [17]. In fact, WA-XRD analysis (**Figure S7** in Supplementary material) showed that after 4 d, P-MBG and P-MBG_1.2Ce showed the characteristic peaks of HA (26° and 31–32° in 2θ); while P-MBG_5.3Ce sample showed a very weak broad peak that could correspond to the presence of a nanocrystalline phase of $CePO_4$.

For B-MBG samples are reported only the FTIR and XRD spectra after 28 d of soaking (**Figure 3**, sections A and B), because for shorter times no significant modification in the spectra were detected. After 28 d of SBF soaking, B-ALG spectrum didn't show any peak in the range 565 and 605 $cm^{-1}$, so the presence of alginate do not interpose in the spectra range of hydroxyapatite. After 28 d, a weak and broad doublet peaks at 605 and 565 $cm^{-1}$ could be noticed in B-MBG and B-MBG_1.2 samples, indicating that a crystalline phosphate phase (hydroxyapatite) was formed. In the other samples (B-MBG_3.6 and B-MBG_5.3) was not possible to detect the formation of bands at 605 and 565 $cm^{-1}$, so the bioactivity was increasingly inhibited with the increase of cerium. Comparing the two series of materials results that the P-samples were more bioactive than the B-samples: this could be due to presence of the polymer that interfere and delay the ionic exchange between the glass and the simulated biological fluid. Furthermore, in each series the most bioactive was the MBG sample and the bioactive performance decreased when increasing the cerium content. As yet reported, the presence of alginate inhibited the formation of hydroxyapatite, in fact the XRD analysis (**Figure 3, section B**) showed the characteristic diffraction maximum of apatite (211) only for B-MBG sample after 28 d of SBF soaking. The presence of a weak peak of $CePO_4$ in the B-MBG5.3Ce sample after 28 d of soaking is probably due to the presence of $CePO_4$ crystal phase in the pristine glass as reported in **Figure 3** section B.

The lower bioactivity of beads with respect the powders revealed by FTIR and XRD analysis is probably due to low surface area where HA could formed, in fact only on the external bead surface the HA can be deposited. In **Figure 4** where reported the micrographs of beads sample before and after 28 d of SBF soaking. After the soaking on the sample surfaces it is possible to detect the formation of a new phase layer. In the B-MBG sample the formed new layer is uniform, while in the samples B-MBG_1.2Ce, B-MBG_3.6Ce and B-MBG_5.3Ce only the new layer did not cover uniformly the original glass surface. The EDS analysis performed in correspondence of these new formed phases revealed a Ca/P ratio of 1.69, 1.75, 1.58 and 1.50 for B-MBG, B-MBG_1.2Ce, B-MBG_3.6Ce and B-MBG_5.3Ce respectively. The theoretical Ca/P ratio in the HA crystal phase is 1.67, we can conclude that the new phase formed on B-MBG sample is compositionally equivalent to HA while the other new phases differ in composition by the HA as a function of cerium content in the glass. This behavior can be explained by the formation of a layer that don not contain only HA but also other phase, in particular Ce-containing phases. In fact in previous work on the glass with a high content of cerium (such as MBG_3.6 and MBG_5.3), during the SBF tests there is the formation of CePO$_4$ on the material surface [16,17]. The Ca/P ratios of 1.58 and 1.50 obtained by EDS are consistent with this behavior.

### 3. 4 Biocompatibility assays

The tests were conducted only on the composite materials (beads) and for convenience the samples are hereafter denominated only with the name of the bioactive glass contained into the bead.

*3.4.1 Cell viability assays.* In **Figure 5A**, MC3T3-E1cells maintained their typical osteoblast morphology in the presence of degradation products from the beads after 1 and 4 d at direct and indirect assays. The results of cell viability were showed in **Figure 5B**, indicating an increase of the proliferation between 1 and 4 d for all the MBG beads. **Figure 5B** shows cell viability after 1 and 4 d at direct and indirect assays in terms of mitochondrial activity. The results show an increase in the levels of proliferation as a function of incubation time on both conditions for MBG beads with and without cerium, showing significant differences from control and cerium MBG beads at indirect assay ($p < 0.05$).

*3.4.2 Cytotoxicity assay. T*o evaluate the cytotoxic effect of the different samples, the amount of lactate dehydrogenase (LDH) released to preosteoblast-like cells cultured in presence of the bead samples was determined after 4 d. The LDH tests (**Figure 6** section (A)) results indicated that the materials weren't cytotoxic since there wasn't significant difference with the control. So, no cytotoxic agent was released compared to the control on MC3T3-E1 cell line.

*3.4.3 Differentiation assay.* ALP a biochemical marker for determining the osteoblast phenotype and is considered as an important factor for assessing bone differentiation and mineralization. **Figure 6** section (B) showed ALP activity of MC3T3-E1 cells on the beads after 7 d of incubation. It could be observed that ALP activity decreased with the amount of cerium, with a significative difference only between the B-MBG_1.2Ce, B-MBG_3.6Ce and B-MBG_5.3Ce samples and the control. These results were coherent with the finding reported in the ref. [45] where the results showed that the materials able to promote the cell proliferation cause a decrement of ALP activity.

*3.4.4 Oxidative stress evaluation*

   *(A) Cell viability in presence of H$_2$O$_2$.* Viability assays in presence of hydrogen peroxide were performed to evaluate the ability of Ce-containing glasses to react against oxidative stress and to improve cell viability. Measurements without hydrogen peroxide were also carried out as a reference viability test. The concentrations of the solutions of H$_2$O$_2$ used in the assays were 400 and 1000 μM, because only in these cases it has been shown a significant difference compared to the control (see **Figure 7** section (A)). Results of the viability assay with 3 beads per well were represented in **Figure 7** section (B) and indicate that the evolution of the viability at H$_2$O$_2$ 400 μM was significantly influenced in a positive way by the amount of cerium contained in the MBGs, demonstrating again the antioxidant properties of this element. In 1000 μM $H_2O_2$ solution the best response was given by B-MBG_5.3Ce. In this case, the cerium-dependence of viability was less evident than in 400 μM $H_2O_2$, and might be due to the quantity of cerium disposable that was not enough to neutralize the amount of $H_2O_2$ in the media.

*(B) Intracellular Reactive Oxygen Species induction assay.* The assay was conducted to qualitatively evaluate the production of the reactive oxygen species by MC3T3-E1 cells, as a result of the induction of oxidative stress by the inoculation of 200 μM hydrogen peroxide solution. The concentration was chosen considering that the objective of the assay was to meanly induce the production of oxidant species and not specifically cells' death. The images reported in **Figure 8** section (A) demonstrated that the control did not show any ROS production. However, with the inoculation of $H_2O_2$ the cell control and MBG beads started to exhibit green spots, indicating that oxidant species have been produced after 1 h. In B-MBG_1.2Ce a significant decrease of ROS species were observed. The two last images represented in **Figure 8** section (B) were obtained to highlight also the positions of the nuclei with DAPI. In B-MBG_3.6Ce and B-MBG_5.3Ce beads a negligible amount of ROS species has been produced. Once more the presence of cerium enhanced the degradation of $H_2O_2$, decreasing the oxidant species in the culture medium and consequently the cellular damage.

## 4. Conclusions

The increment of cerium content in the glass causes a decrease of the textural properties ($S_{BET}$ and $V_P$) of MBG powders. The scaffolds (beads), showed a remarkable drop of $S_{BET}$ and $V_P$ compared to powders, but they still maintain good enough textural properties to be used in bone regeneration applications. The increment of cerium content increases the catalase mimetic activity, both for powders and beads. However, the cerium increment decreases the in vitro bioactivity of the beads in simulated body fluid. The beads did not compromise the cellular proliferation, on the contrary at higher Ce-contents, B-MBG_3.6 and B-MBG5.3, their presence improved the preosteoblast cells proliferation, especially in the indirect assay after 4 d, without cytotoxic effects. Nevertheless, the cell differentiation decrease as a function of Ce-content in the samples. Under oxidative stress induced by the inoculation of $H_2O_2$ the elimination of oxidant species was shown when increasing the amount of cerium in the beads. Therefore, B-MBG_1.2Ce and B-MBG_3.6 Ce beads are excellent candidates as biocompatible scaffolds capable of counteract the oxidative stress.

**Acknowledgements**

The Authors thank Paola Luches for XPS measurements. This study was supported by research grants from Instituto de Salud Carlos III (PI15/00978) project co-financed with the European Union FEDER funds, the European Research Council (ERC-2015-AdG) Advanced Grant Verdi-Proposal No. 694160 and MINECO MAT2015-64831-R project.

**Conflicts of Interest:** The authors declare no conflict of interest.

**Appendix A. Supplementary material**

# SCHEME CAPTION

**Scheme 1** Schematic preparation of the alginate/glass beads.

# FIGURE CAPTIONS

**Figure 1 (A):** Solid-state $^{29}$Si single-pulse MAS-NMR spectra of P-MBG, P-MBG_1.2Ce, P-MBG_3.6Ce, P-MBG_5.3Ce samples (---experimental spectra, ---fitting reconstruction and ---- $Q^n$ contributions). The areas for the $Q^n$ units were calculated by Gaussian line-shape deconvolution and are displayed in green. **(B):** Chemical shifts (CS), relative % population of $Q^n$ species and mean $<Q^n>$ value.

**Figure 2** SEM micrographs of the surface of B-MBG and B-MBG_3.6Ce and of the cross-section of B-MBG_1.2Ce and B-MBG_5.3Ce.

**Figure 3 (A):** FTIR-ATR spectra of B-MBG, B-MBG_1.2Ce, B-MBG_3.6Ce, B-MBG_5.3Ce and B-ALG after 28 d of soaking in SBF (dotted lines = characteristic bands of HA at 605 and 565 cm$^{-1}$). **(B):** XRD patterns of B-MBG, B-MBG_1.2Ce, B-MBG_3.6Ce, B-MBG_5.3Ce after 28 d in SBF.

**Figure 4** SEM micrographs and EDS spectra of B-MBG, B-MBG_1.2Ce, B-MBG_3.6Ce and B-MBG_5.3Ce before and after 28 d of soaking in SBF.

**Figure 5 (A):** Morphological evaluation of cell viability of preosteoblast cells after 1 d of culture using indirect and direct assays. **(B):** Cell viability (Alamar Blue) of preosteoblastic cells after 1 and 4 d of culture. (* = significant differences between control and samples after 4 d, $p < 0.05$).

**Figure 6 (A):** LDH release into the medium from MC3T3-E1 cells after 4 d of culture using indirect and direct assays. Values shown are means and the standard errors were included. **(B):** ALP activity of MC3T3-E1 cells cultured 7 d on beads containing bioactive glasses. Values shown are means and the standard errors were included. ($\$$= significant differences between control/B-MBG and the samples B-MBG_1.2Ce, B-MBG_3.6Ce and MBG_5.3Ce, $p < 0.05$).

**Figure 7 (A):** Cell viability (MTS assay) of MC3T3-E1 cells after 6 h of incubation with different concentrations of $H_2O_2$. The values shown are means with the standard errors. **(B):** Cell viability of preosteoblast cells after 6 h of culture in the presence of beads with 400 and 1000 µM $H_2O_2$ solutions. First column of each series represented a control without $H_2O_2$. The values shown are means with standard errors. (*significant differences between control and the samples; $ between B-MBG and B-MBG_1.2Ce, MBG_3.6Ce and B-MBG_5.3Ce; # between control, B-MBG and B-MBG_1.2Ce, B-MBG_3.6Ce, B-MBG_5.3Ce; & between B-MBG_5.3Ce and the other samples and the control, $p < 0.05$).

**Figure 8. (A):** Confocal microscope images of preosteoblast cells after 1 h of incubation with hydrogen peroxide (200 µM). The oxidation stress was proportional to the green spots. The green spots are due to the oxidation of 2',7'-dichlorfluorescein diacetate (DCFH-DA), a dye that became fluorescent (green) after oxidation operate by the ROS. **(B):** In B-MBG_3.6 and B-MBG_5.3 the nuclei of vital cells are marked with DAPI (blue).

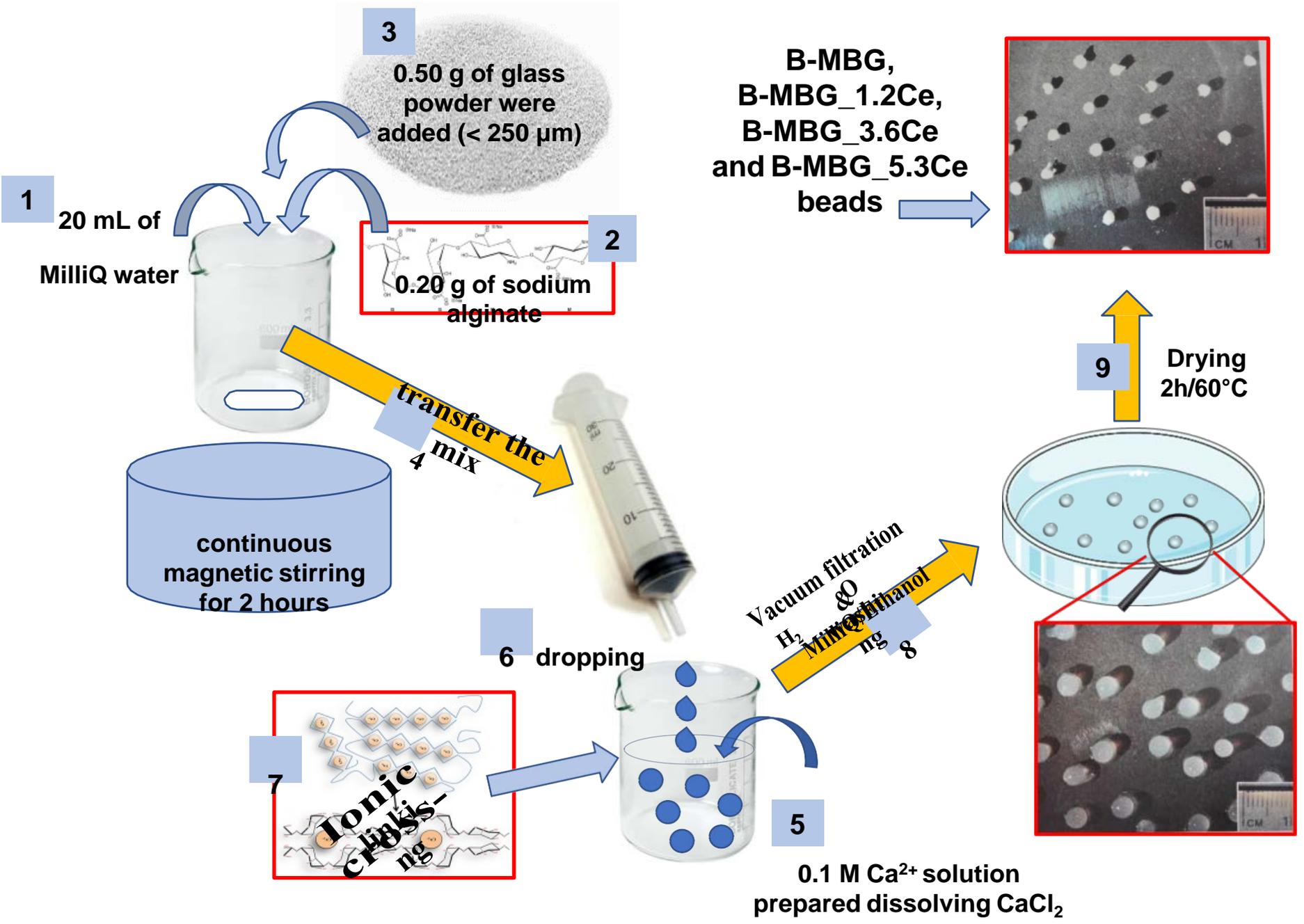

in 30 mL MilliQ water

## $^{29}$Si Single Pulse MAS-NMR

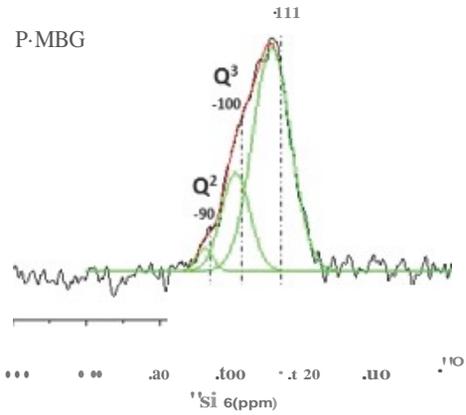
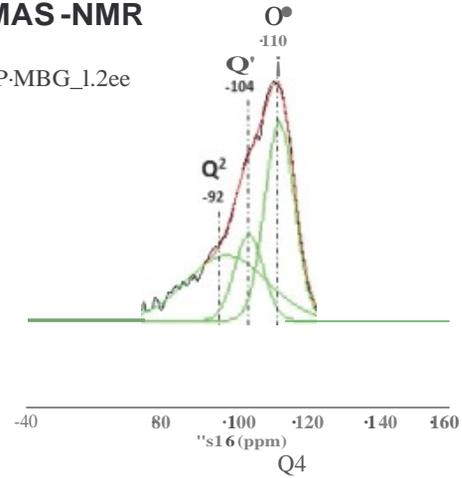
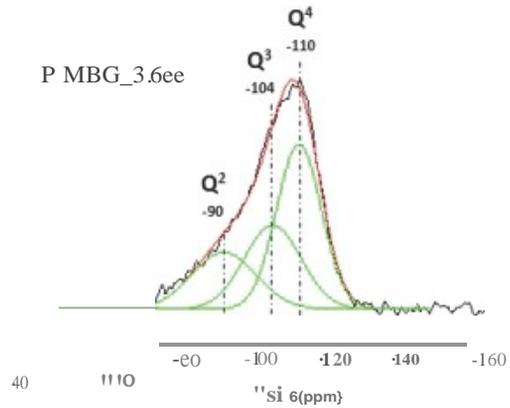
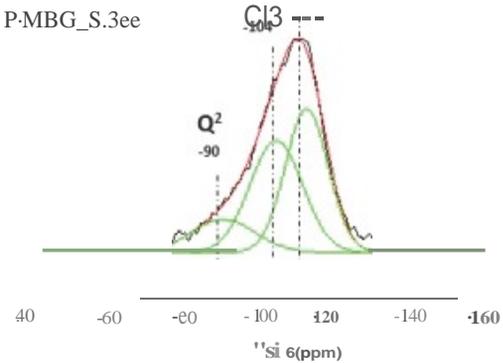

(A)

(B)

| Sample | $^{29}$Si | | | | | | $\langle Q^n \rangle$ |
|---|---|---|---|---|---|---|---|
| | Q4 | | Q3 | | Q2 | | |
| | δs (ppm) | Area (%) | δs (ppm) | Area (%) | δs (ppm) | Area (%) | |
| P MBG | -111 | 72.9 | -100 | 24.0 | -90 | 3.1 | 3.70 |
| P MBG_1.2Ce | -111 | 44.1 | -104 | 34.4 | -92 | 21.5 | 3.23 |
| P MBG_3.6Ce | -110 | 46.8 | -104 | 30.1 | -90 | 23.0 | 3.24 |
| P MBG_5.3Ce | -110 | 43.6 | -104 | 41.7 | -90 | 14.8 | 3.29 |

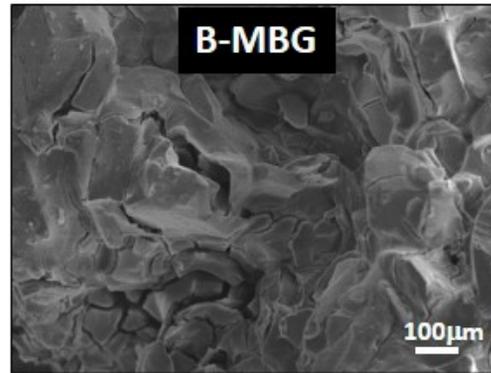 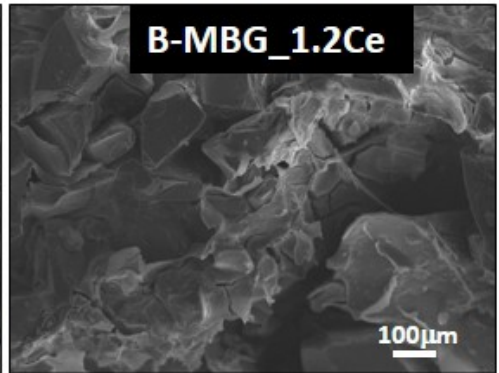
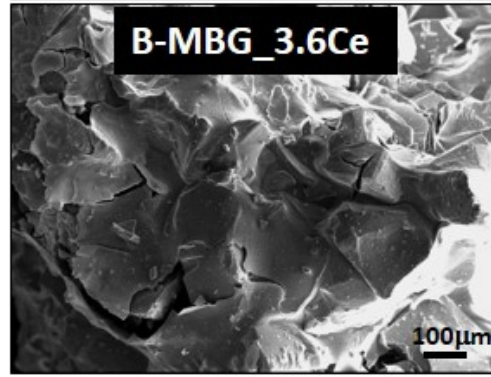 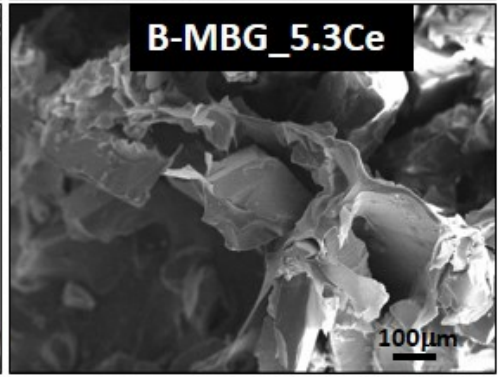

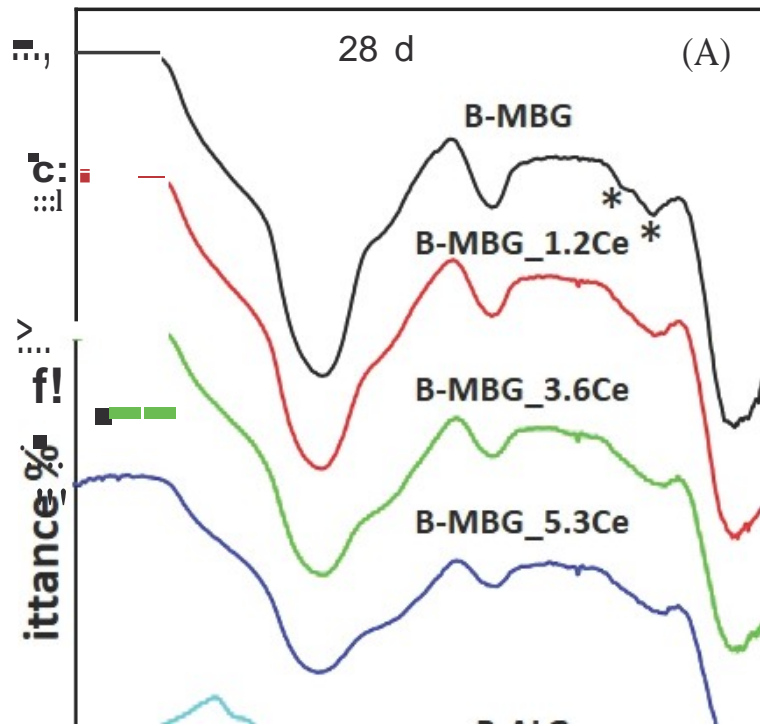
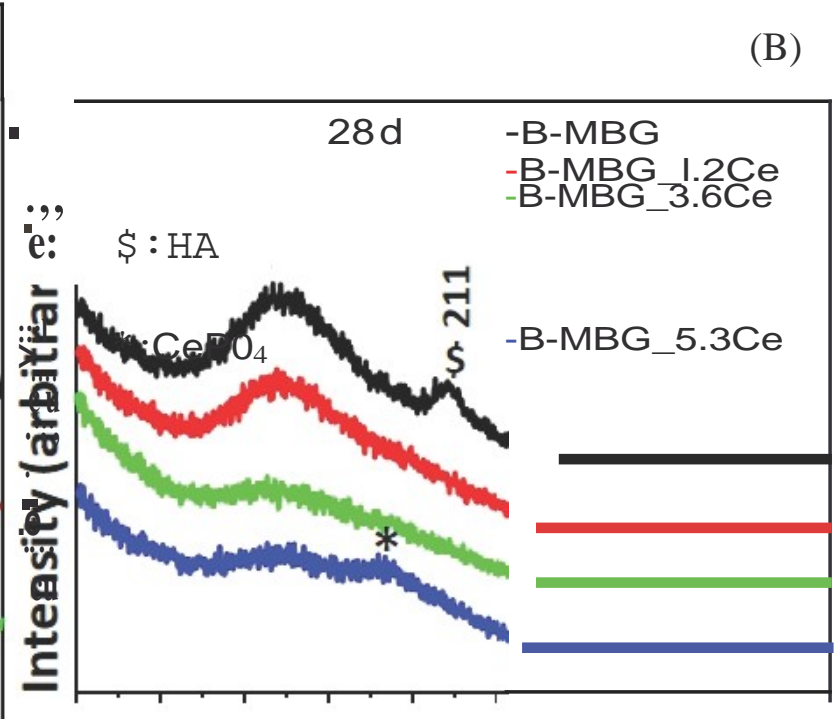

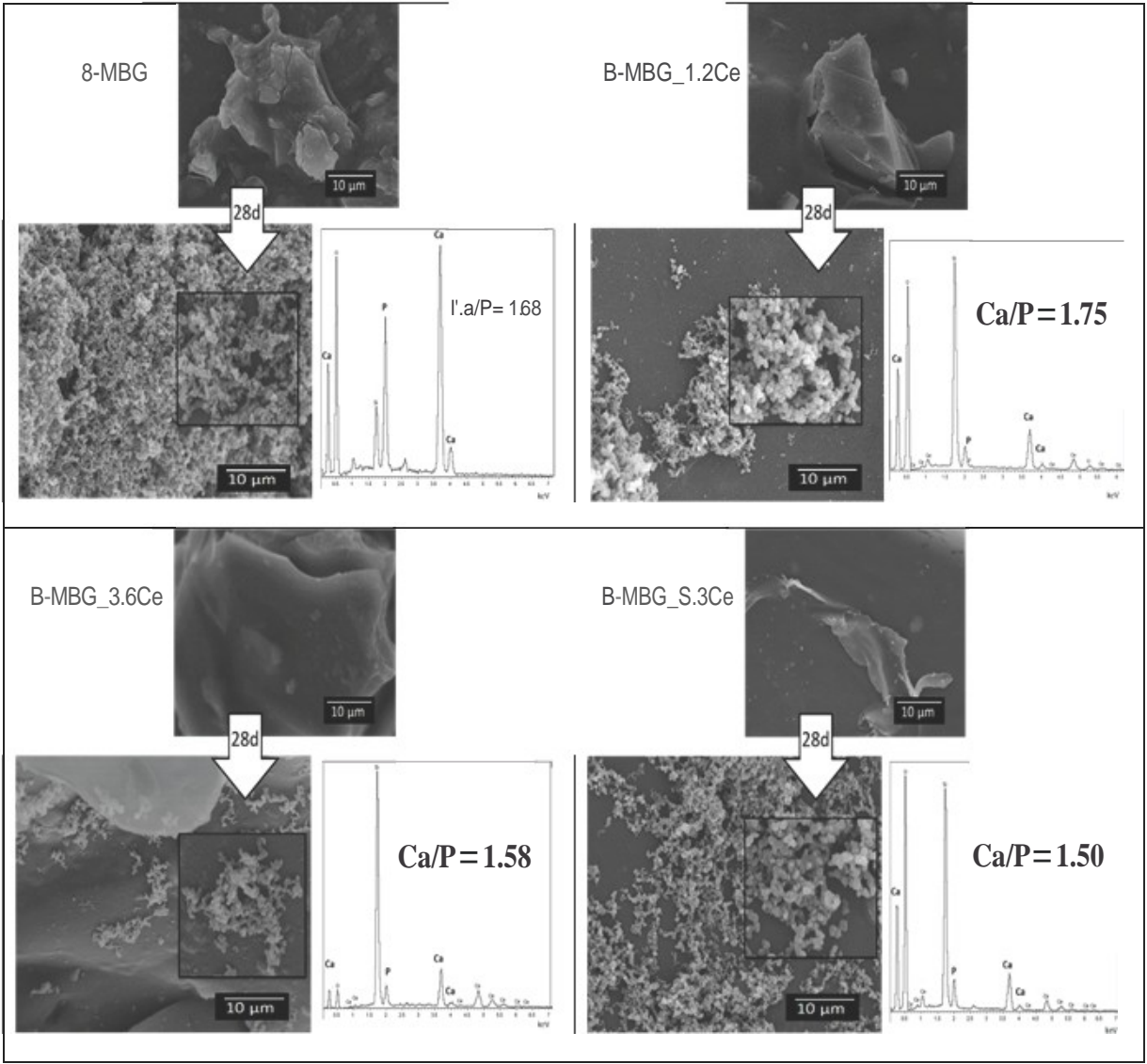

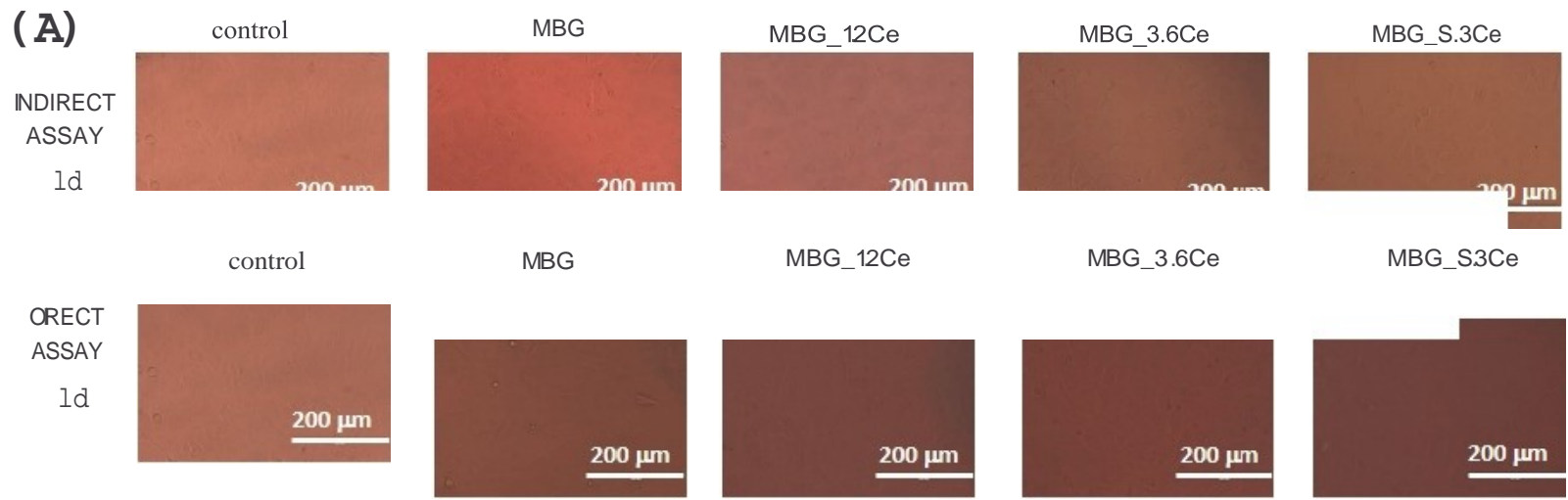
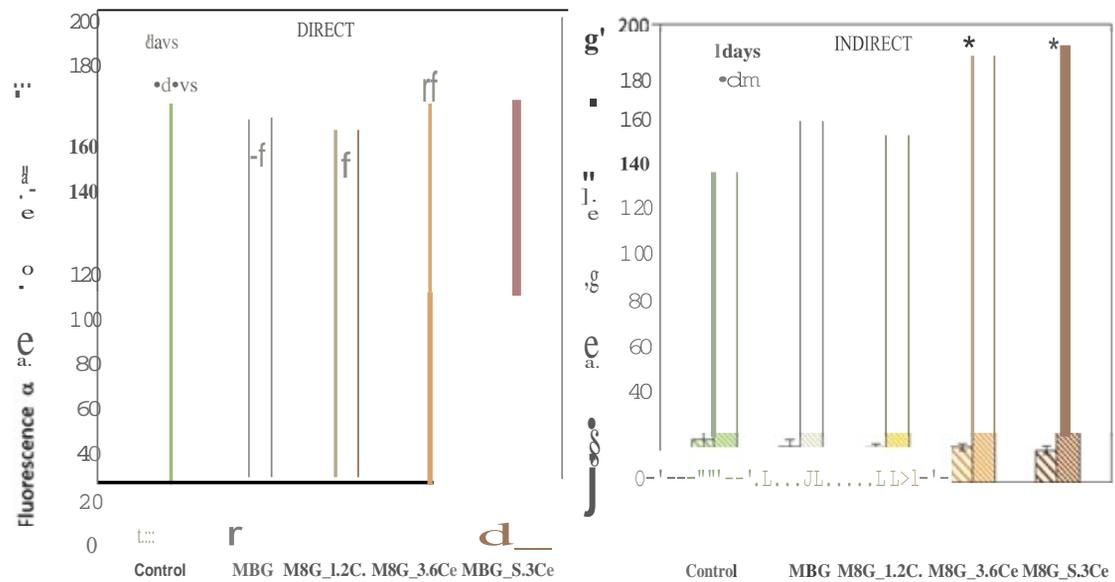

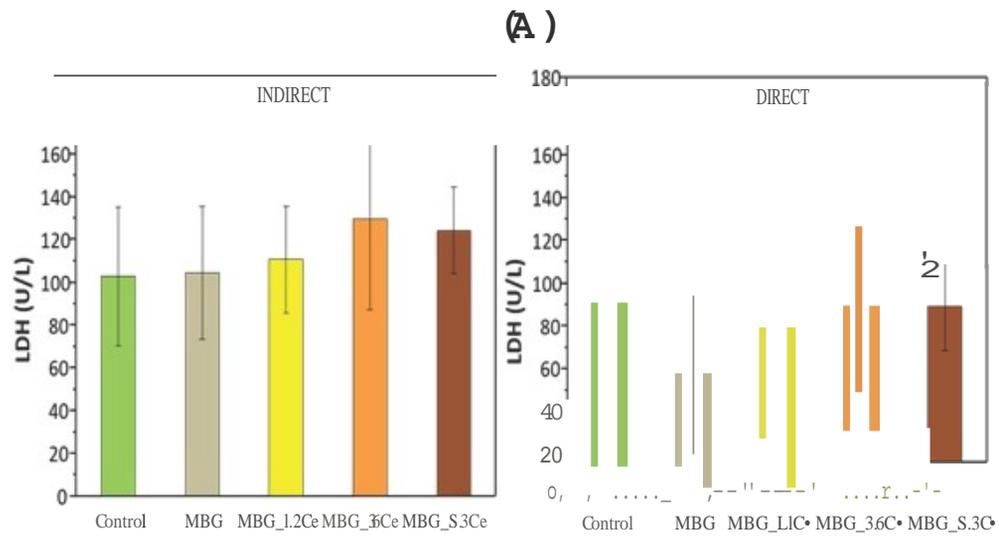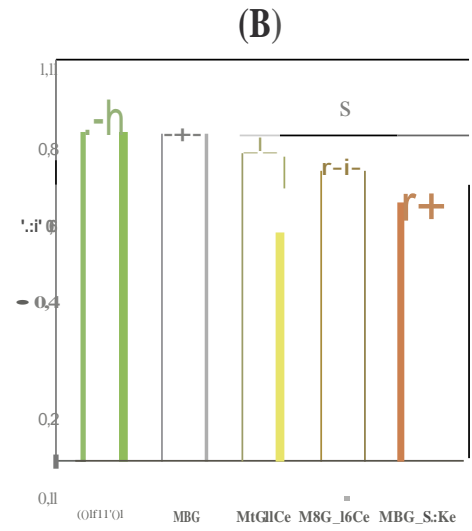

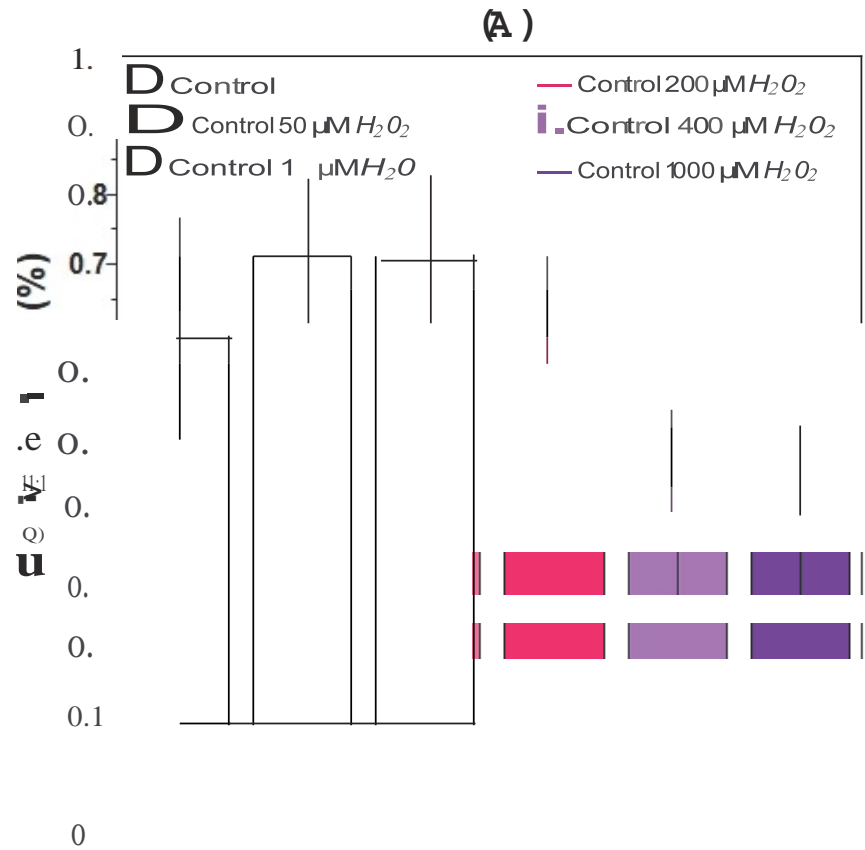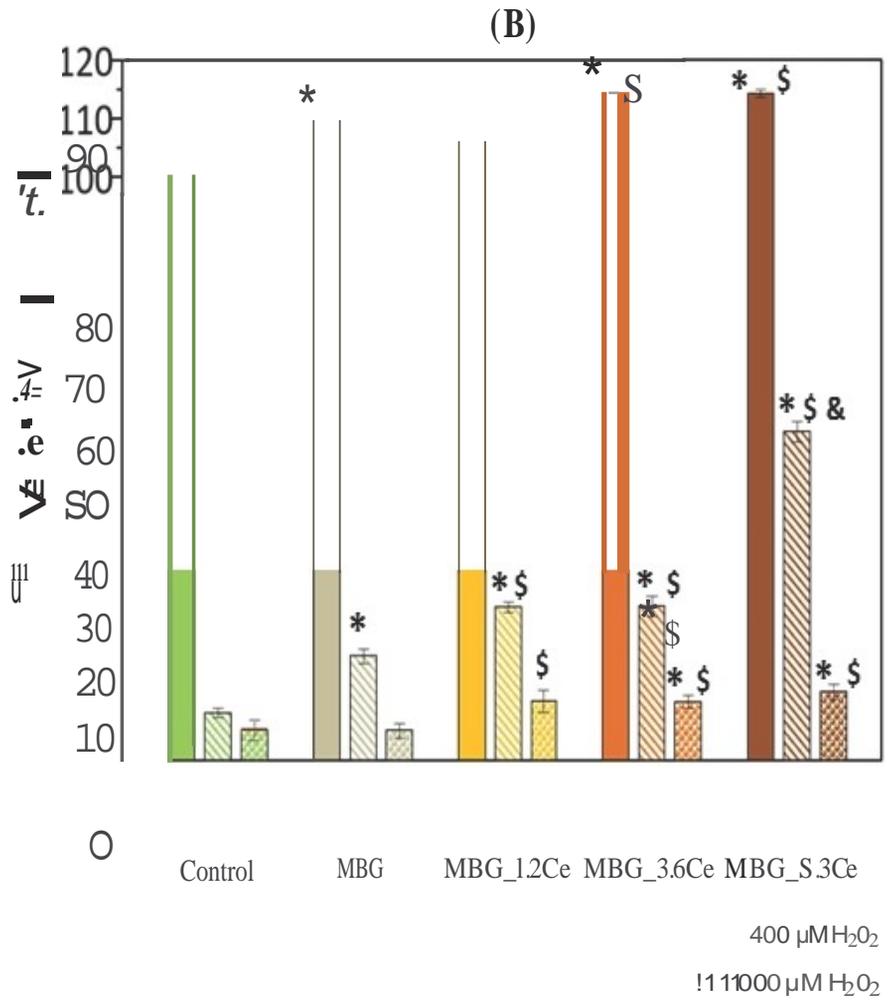

(A)

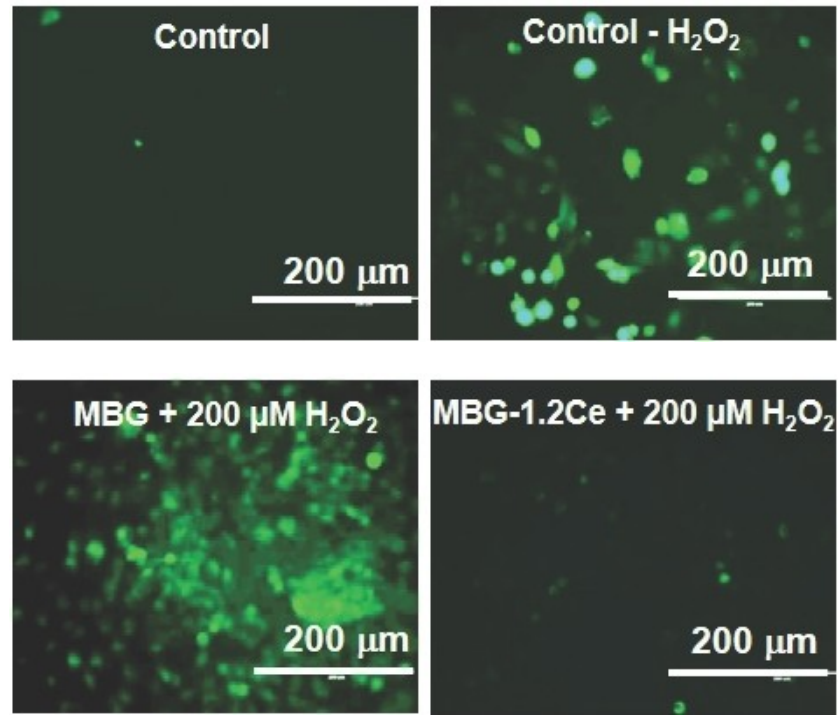

(B)

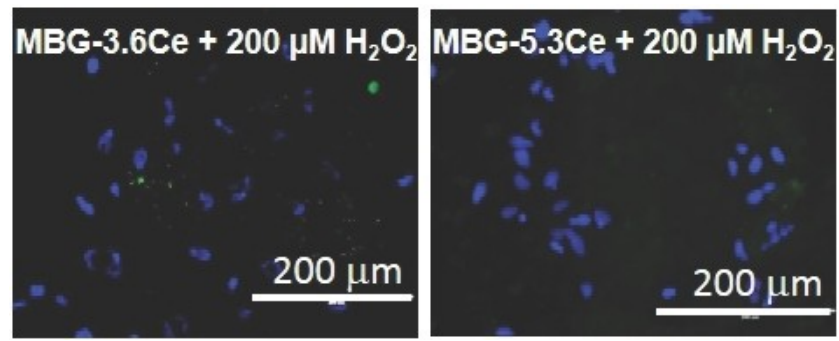

**Table 1**. Theoretical (and determined by XRF ± std. dev.) molar compositions of synthesized mesoporous bioactive glasses (MBG).

| Glass | $SiO_2$ (mol%) | CaO (mol%) | $P_2O_5$ (mol%) | $CeO_2$* (mol%) |
|---|---|---|---|---|
| **MBG** | 80.0 | 15.0 | 5.0 | - |
|  | (79.3 ± 0.2) | (16.4 ± 0.2) | (4.3 ± 0.1) |  |
| **MBG_1.2Ce** | 79.1 | 14.8 | 4.9 | 1.2 |
|  | (79.0 ± 0.5) | (15.2 ± 0.5) | (4.2 ± 0.1) | (1.6 ± 0.2) |
| **MBG_3.6Ce** | 77.1 | 14.5 | 4.8 | 3.6 |
|  | (76.5 ± 0.3) | (15.1 ± 0.1) | (4.1 ± 0.1) | (4.3 ± 0.2) |
| **MBG_5.3Ce** | 75.8 | 14.2 | 4.7 | 5.3 |
|  | (75.4 ± 0.4) | (14.3 ± 0.1) | (4.1 ± 0.1) | (6.2 ± 0.4) |

*The amount of cerium was expressed as $Ce(IV)O_2$

**Table 2.** Specific surface area ($S_{BET}$), mean pore diameter ($D_P$) and pore volume ($V_P$) of P-MBGs and B-MBGs samples.

| Sample | $S_{BET}$ (m²/g) | $D_P$ (nm) | $V_P$ (cm³/g) |
|---|---|---|---|
| **P-MBG** | 405 | 4.6 | 0.43 |
| **P-MBG_1.2Ce** | 374 | 3.5 | 0.32 |
| **P-MBG_3.6Ce** | 324 | 3.3 | 0.24 |
| **P-MBG_5.3Ce** | 311 | 2.9 | 0.23 |
| **B-MBG** | 173 | 3.9 | 0.17 |
| **B-MBG_1.2Ce** | 150 | 3.5 | 0.13 |
| **B-MBG_3.6Ce** | 136 | 3.3 | 0.12 |
| **B-MBG_5.3Ce** | 94 | 3.1 | 0.07 |

**Table 3** $H_2O_2$ molar concentration after soaking P-MBGs and B-MBGs samples. Data are the mean of three measurements. Standard deviation (± std. dev.) is always ≤ 0.02 M.

| Samples | Time (h) | | | | | | | | |
|---|---|---|---|---|---|---|---|---|---|
| | 0 | 1 | 2 | 4 | 8 | 24 | 48 | 96 | 168 |
| **P-MBG** | 1.00 | 0.92 | 0.93 | 0.92 | 0.94 | 0.94 | 0.89 | 0.87 | 0.85 |
| **P-MBG_1.2Ce** | 1.00 | 0.92 | 0.92 | 0.90 | 0.90 | 0.95 | 0.75 | 0.56 | 0.27 |
| **P-MBG_3.6Ce** | 1.00 | 0.93 | 0.92 | 0.90 | 0.89 | 0.86 | 0.56 | 0.28 | 0.06 |
| **P-MBG_5.3Ce** | 1.00 | 0.92 | 0.90 | 0.86 | 0.76 | 0.38 | 0.05 | 0.00 | 0.00 |
| **B-MBG** | 1.00 | 0.94 | 0.92 | 0.92 | 0.91 | 0.87 | 0.84 | 0.82 | 0.77 |
| **B-MBG_1.2Ce** | 1.00 | 0.93 | 0.92 | 0.91 | 0.87 | 0.95 | 0.70 | 0.48 | 0.21 |
| **B-MBG_3.6Ce** | 1.00 | 0.93 | 0.91 | 0.88 | 0.83 | 0.80 | 0.48 | 0.19 | 0.00 |
| **B-MBG_5.3Ce** | 1.00 | 0.91 | 0.91 | 0.80 | 0.68 | 0.31 | 0.01 | 0.00 | 0.00 |

**Conflicts of Interest:**

**Cerium (III) and (IV) containing mesoporous glasses/alginate beads for bone regeneration: bioactivity, biocompatibility and reactive oxygen species activity**

E. Varini[1,§], S. Sánchez-Salcedo[2,3,§], G. Malavasi[1], G. Lusvardi[1], M. Vallet-Regí[2,3], A.J. Salinas[2,3,*]

[1] Dpt. Chemical and Geological Sciences, University of Modena and Reggio Emilia, Via G. Campi 103, 41125 Modena, Italy.

[2] Dpt. Química en Ciencias Farmacéuticas, Facultad de Farmacia, Universidad Complutense de Madrid; Instituto de Investigación Hospital 12 de Octubre, imas12, 28040 Madrid, Spain.

[3] Networking Research Center on Bioengineering, Biomaterials and Nanomedicine (CIBER-BBN) Madrid, Spain

The authors declare no conflict of interest.